\documentclass[jcp,aps,showpacs,superscriptaddress,twocolumn,a4paper]{revtex4}
\usepackage{amsmath}
\usepackage{amsfonts}
\usepackage{amssymb}
\usepackage{graphicx}
\usepackage{color}

\def\pj{\phi_{J}^{1}}
\def\pe{\phi_J'}
\def\p1{\phi_{\rm Vor}}
\def\Ze{Z_{\rm Euler}}
\def\Ziso{Z_{\rm iso}}

\def\<{\langle}
\def\>{\rangle}

\def\E0{E_{\rm T=0}}

\begin{document}
\title{The first jamming crossover: geometric and mechanical features}

\author{Massimo Pica Ciamarra}\email[]{massimo.picaciamarra@spin.cnr.it}
\affiliation{
CNR--SPIN, Dipartimento di Scienze Fisiche,
Universit\`a di Napoli Federico II, I-80126, Napoli, Italy
}
\author{Peter Sollich}
\affiliation{
King’s College London, Department of Mathematics, Strand, London WC2R 2LS, United Kingdom
}

\date{\today}
\begin{abstract}
The jamming transition characterizes athermal systems of particles interacting via finite range repulsive potentials,
and occurs on increasing the density when particles cannot avoid making contacts with those of their first coordination shell.
We have recently shown [M. Pica Ciamarra and P. Sollich, arXiv:1209.3334] that the same systems are also characterized by a series of jamming
crossovers. These occur at higher volume fractions as particles are forced to make contact with those of
subsequent coordination shells. 
At finite temperature, the crossovers give rise to dynamic and thermodynamic density anomalies, 
including a diffusivity anomaly and a negative thermal expansion coefficient. 
Density anomalies may therefore be related to structural changes occurring at the jamming crossovers.
Here we elucidate these structural changes, investigating the evolution of the structure and of the mechanical properties 
of a jammed system as its volume fraction varies from the jamming
transition to and beyond the first jamming crossover.
We show that the first jamming crossover occurs at a well defined volume fraction, and that
it induces a rearrangement of the force network causing a softening of the system. It also causes
qualitative changes in the normal mode density of states and the spatial properties of the normal mode vectors.
\end{abstract}
\pacs{61.43.Er; 62.10.+s; 61.20.Ja}
\maketitle

\section{Introduction}
Athermal soft matter systems with particles interacting
via repulsive contact forces,
such as granular materials, foams and emulsions, 
undergo a percolating transition known as the jamming transition
when, on increasing
the density, particles can no longer avoid making contacts. We refer to Refs.~\cite{vanHecke2010,Liu2010} for recent reviews.
At the jamming transition a percolating network of contact forces
appears, and the system acquires mechanical rigidity.
We have recently shown that the jamming transition is
the first of a series of higher--order jamming crossovers~\cite{toappear}.
The origin of these crossovers can be understood by considering
the radial distribution function of dense particle systems, which allows one to define
a sequence of coordination shells. The jamming transition marks the point where each particle
contacts those of its first coordination shell. The jamming crossovers then
occur at higher density as particles interact with those of subsequent coordination shells.
Accordingly, at zero temperature these crossovers are characterized by an increase of the mean number of contacts per particle.
This is accompanied by a change in the mechanical response of the
system, which becomes more non-affine. At finite temperature
this leads to density anomalies, such as an increasing diffusivity upon isothermal compression, and a negative thermal expansion coefficient~\cite{toappear}.
These anomalies are analogous to those observed in water and in some other network forming liquids~\cite{Jagla,Stillinger1997,Mausbach,Stanley2005}.
Accordingly, the investigation of the jamming crossovers may shed light on the microscopic origin of these phenomena.
In particular, the observed correlation between the jamming crossovers and the density anomalies suggest that these
anomalies might be related to the structural changes occurring at the crossovers.

In this paper we investigate the structural changes
occurring at the first jamming crossover in a bidisperse mixture
of disks interacting via a finite-range, repulsive harmonic potential.
We show that there is a well defined volume fraction $\pe$ at which 
the formation of contacts with particles of the first coordination shell ends,
and a larger well defined volume fraction $\pj > \pe$ at which the formation
of contacts with particles of the second coordination shell begins; this is the location of 
the first jamming crossover.
We provide evidence that in the volume fraction range $(\pe,\pj)$ two particles are in contact
if and only if they are also Voronoi neighbors, and the average contact number per particle
$Z$ is then constant and fixed to $Z = \Ze = 6$ by Euler's theorem for planar graphs.
The formation of contacts between non--Voronoi neighbors, which occurs for $\phi > \pj$,
induces mechanical instabilities and avalanches that restructure the whole force network,
with the net effect of producing a large number of weakly compressed bonds.
These structural changes allow us to rationalize the softening of the system on compression,
as well as the volume fraction dependence of the normal mode density of states and the spatial structure of the normal mode vectors.

\section{Numerical model}
We investigated the zero temperature properties of a $50$:$50$ 
mixture of soft disks, with diameters $D_{\rm l}$ and $D_{\rm s} = D_{\rm l}/1.4$, and equal mass $M$.
Two particles $i$ and $j$ interact via a harmonic potential when in contact,
\begin{equation}
 V(r_{ij}) = \frac{1}{2}k \left( \frac{\delta_{ij}}{D_l} \right)^2 \theta(\delta_{ij}),
\label{eq:potential}
\end{equation}
where $\delta_{ij} = D_{ij} - r_{ij}$ is the overlap between the
particles. This is expressed in terms of $D_{ij}=(D_i+D_J)/2$, the
average of the two particle diameters $D_i$ and $D_j$, and the
distance $r_{ij} = |\bf{r}_i-\bf{r}_j|$ between the particles; $\theta(x)$
in Eq.~(\ref{eq:potential}) is the Heaviside function.
We will take $M$, $D_{\rm l}$ and $k$ as our units of mass, length and energy, respectively.
For each volume fraction considered, $50$ independent jammed configurations were generated by minimizing
the energy via the conjugate--gradient algorithm~\cite{Ohern}, starting from an initial  configuration with randomly placed particles.
We considered values of volume
fraction separated by at most $\Delta \phi = 5 \times 10^{-2}$, and used measurements taken at successive
values of the volume fraction to estimate volume fraction derivatives~\cite{note}.
The results reported below concern systems of $N = 10^4$ particles unless stated otherwise.

\begin{figure}
\includegraphics*[scale=0.35]{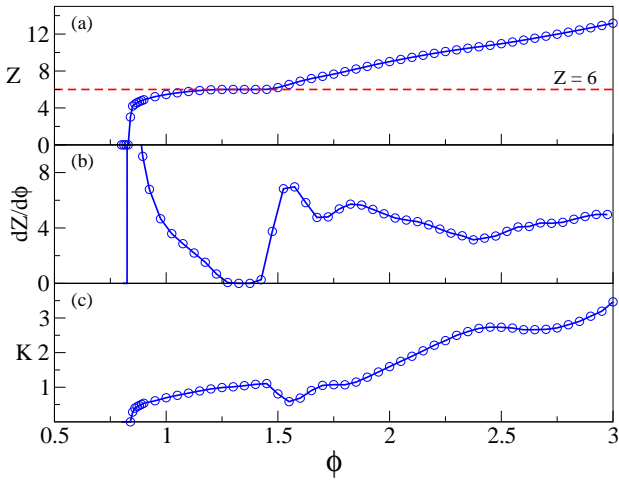}
\caption{\label{fig:crossovers} 
(color online) Volume fraction dependence of the mean contact number (a), its volume
fraction derivative (b), and the bulk modulus (c).}
\end{figure}

\section{Jamming crossovers}
Fig.~\ref{fig:crossovers}c illustrates the volume fraction dependence
of the mean contact number per particle $Z$, its volume fraction derivative $dZ/d\phi$,
and the bulk modulus $K$. 
In agreement with previous results for harmonic disks~\cite{Ohern,vanHecke2010}, 
at the jamming transition, $\phi_J \simeq 0.84$, the mean contact
number jumps to the isostatic value $\Ziso = 2d = 4$, and $dZ/d\phi$ diverges. 
In addition, on approaching the jamming transition from above,
$Z$ varies as $Z-\Ziso \propto (\phi-\phi_J)^{0.5}$. The bulk modulus, finally, changes discontinuously at the transition.

Here we focus on the volume fraction dependence of these quantities at higher volume fractions, and note
that while $Z$ grows monotonically on compression (Fig.~\ref{fig:crossovers}a), the
rate of formation of new contacts oscillates (Fig.~\ref{fig:crossovers}b). 
These oscillations can be interpreted as higher order jamming crossovers.
Indeed, while at the jamming transition $dZ/d\phi$ diverges when
particles make contacts with other particles in their first coordination shell, at higher volume fraction $dZ/d\phi$ increases again as 
particles form contacts with subsequent coordination shells. 
The oscillations of $dZ/d\phi$ are correlated with those
of the bulk modulus $K$, illustrated in Fig.~\ref{fig:crossovers}c,
suggesting a relation between the geometric and the mechanical properties
of the system. In particular, the jamming crossovers appear to induce a softening of the system,
as $K$ decreases when $dZ/d\phi$ increases.
The finite temperature counterparts of this softening include diffusivity anomalies, and a negative thermal expansion coefficient~\cite{toappear}. 

\section{The first crossover: geometric properties}
\subsection{The crossover volume fraction, $\pj$}
\begin{figure}
\includegraphics*[scale=0.33]{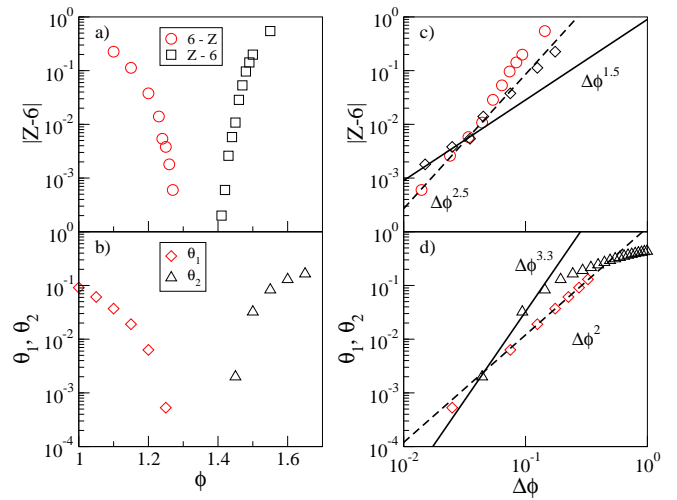}
\caption{\label{fig:op} 
(color online) Dependence of $|Z-\Ze|$, with $\Ze = 6$,
on the volume fraction (panel a) and on $\Delta \phi$ (panel b), where 
$\Delta\phi = \pe - \phi$ for $\phi < \pe$, and
$\Delta\phi = \phi-\pj$ for $\phi > \pj$. 
Dependence of $\theta_1$ and $\theta_2$ (see text) on $\phi$ (panel c) and
on $\Delta \phi$ (panel d). The figure clarifies that $Z=\Ze$ in the range $\pe < \phi < \pj$,
as two particles are in contact if and only if they are Voronoi neighbors.}
\end{figure}

In the affine approximation, the rate of formation of new contacts is simply related to the radial distribution
function. Each jamming crossover would then be identified with the
volume fraction where the formation of contacts with particles of a
new shell starts. It is possible that the process of formation of such
contacts might terminate before the next jamming crossover begins.
In the intervening range of volume fractions, $Z$ is then constant and $dZ/d\phi = 0$:
all contacts with particles of a given shell are established, and no
contacts with particles of the subsequent shell form. However, this scenario is not very common
as successive crossovers usually mix, leading to a rate of formation of new contacts that is generically nonzero, $dZ/d\phi > 0$.

Harmonic disks in two dimensions are peculiar in this respect, as the formation of contacts in the first neighbor shell, which starts at the jamming transition $\phi_J$, ends at a volume fraction $\pe$
well below the first jamming crossover at $\pj > \pe$.
We show that this is the case by investigating the volume dependence of $|Z-6|$, 
motivated by the fact that Fig.~\ref{fig:crossovers}a suggests the existence of a volume fraction range where the number of contacts is constant at $Z = 6$.
Fig.~\ref{fig:op}a shows that $|Z-6|$ vanishes in a finite volume fraction range, which extends
from $\pe$ to $\pj$. We estimate the two volume fractions delineating this regime by assuming a power-law dependence 
of $|Z-6|$ on $\Delta\phi$. The latter is defined as $\Delta \phi = \pe-\phi$ when $Z < 6$, and as $\Delta \phi = \phi-\pj$ when $Z > 6$.
From the numerical fits in Fig.~\ref{fig:op}c we determine $\pe \simeq 1.27$ and $\pj \simeq 1.40$. We note,
however, that accurate estimates of these volume fractions are difficult to obtain because of the high exponents characterizing the relevant power laws.
In addition, it is worth remembering that the values of $\pe$ and $\pj$ may be slightly protocol dependent,
in analogy with the location $\phi_J$ of the conventional jamming transition~\cite{protocol}.

The value $Z = 6$ of the mean contact number in the volume fraction range leading up to the first jamming crossover can be rationalized.
Indeed, Euler's theorem for planar graphs fixes to $6$ the average connectivity of any tessellation of space in two dimensions,
in the limit of large system size. Since the simplest tessellation is the Voronoi one, we speculate that $Z = \Ze = 6$
for $\pe < \phi < \pj$ because in this volume fraction range two particles are in contact if and only if they are also Voronoi neighbors. 
To check this hypothesis, we investigate
the volume fraction dependence of two order parameters related to the Voronoi (radical~\cite{Radical}) tessellation.

The first one, $\theta_1$, is the fraction of Voronoi neighbors that are not in contact;
the second one, $\theta_2$, is the fraction of contacts that occur between particles which are not Voronoi neighbors.
Fig.~\ref{fig:op}b shows that $\theta_1$ and $\theta_2$ 
are non-zero for $Z < \Ze$ and for $Z > \Ze$, and vanish 
at $\pe$ and at $\pj$, respectively. Indeed, their dependence
on $\Delta \phi$ is well described by power laws, as in Fig.~\ref{fig:op}d.
The results of Fig.~\ref{fig:op}b confirm that, for $\pe < \phi < \pj$, 
two particles are in contact if and only if they are also Voronoi neighbors, and thus rationalize the observed constancy of the mean contact number $Z$ in this volume fraction range. We note as an aside that
this result is not observed for all interaction potentials~\cite{toappear}, and does not hold in three dimensions where Euler's theorem does not constrain the mean contact number of the Voronoi tessellation.

\subsection{Overlap PDF}
\begin{figure}[t]
\includegraphics*[scale=0.33]{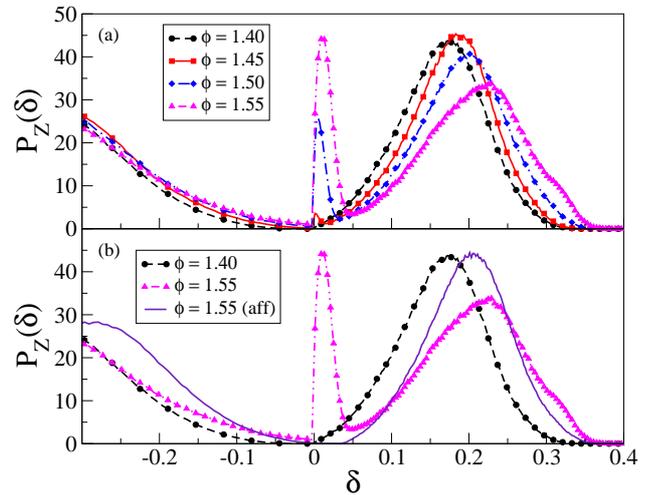}
\caption{\label{fig:Pd} (color online) 
(a) Overlap probability distribution $P_Z(\delta)$, for increasing volume fractions beyond
the first jamming crossover, as indicated. A negative $\delta$ corresponds to the distance between the surfaces of non--interacting particles.
The distribution is normalized across the positive overlaps, so that $\int_0^1P_Z(\delta)d\delta = Z$.
(b) To highlight the non--affine rearrangements occurring at the jamming crossovers,
we plot the actual overlap probability distribution at volume fraction $\phi = 1.55$
and the one resulting from an affine compression of the system from $\phi = 1.40$ to
$\phi = 1.55$ (solid line).
}
\end{figure}

We next look at what happens for volume fractions above $\pj$, where the mean contact number $Z$ begins to rise again with $\phi$.
Particles now start to make contacts with those
of their second coordination shell. As these new contacts are formed, 
the entire force network undergoes a marked restructuring. To 
investigate the nature of this, we study the volume fraction evolution of the overlap probability distribution function, $P_Z(\delta)$.
We normalize $P_Z$ across the interacting particles, so that $\int_0^1 P_Z(\delta) d \delta = Z$. 
A negative overlap corresponds to the distance between the surfaces of two particles that are not in contact.
(This is the reason why we monitor distributions of overlaps rather than forces; while for positive $\delta$ force and overlap are identical for our harmonic potential, forces are zero for all particle pairs with negative $\delta$ and so provide less information on the packing geometry.)
Figure~\ref{fig:Pd}a shows that the evolution of $P_Z(\delta)$ across the first jamming crossover
is characterized by the growth of a peak at small values of $\delta > 0$, and by the
simultaneous decrease of the peak at larger overlaps. 

Such a sharp variation of the overlap probability distribution signals the presence
of a non--affine evolution of the force network on compression. Indeed, in the affine approximation
the overlap $\delta_{ij}$ between particles $i$ and $j$ evolves as (with the spatial dimension being $d=2$ in our case)
\[
\frac{\partial\delta_{ij}}{\partial\phi} = \frac{1}{d\phi}\left(D_{ij}-\delta_{ij}\right),
\]
and consequently the overlap distribution around $\delta=0$ mainly shifts on compression.
We illustrate this point in Fig.~\ref{fig:Pd}b, where we compare the distributions $P_Z$
at volume fraction $\phi = 1.55$ for two different systems. One is obtained
by energy minimization via our usual conjugate--gradient protocol, the other
by minimizing the energy at $\phi = 1.40$ and then affinely compressing the system to $\phi = 1.55$.
The two distributions differ strongly, and the comparison demonstrates
that the jamming crossover induces a non--affine transformation of the force network on compression whose main effect is a large increase in the number of small overlaps. 

\begin{figure}[t!!]
\includegraphics*[scale=0.4]{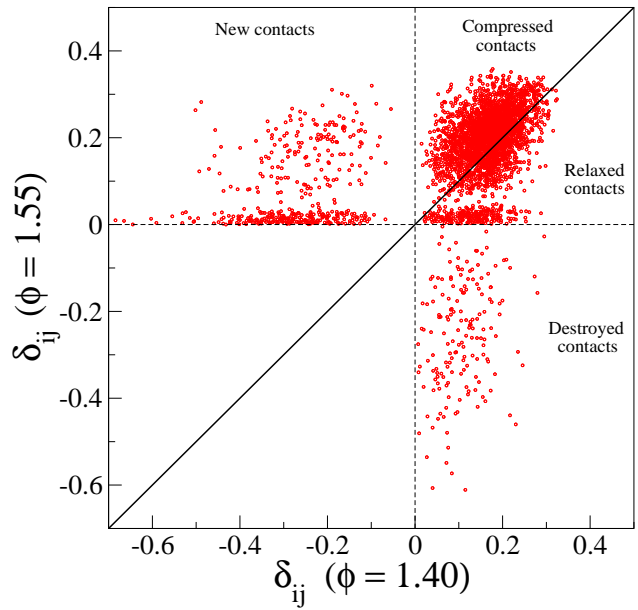}
\caption{\label{fig:scatter} 
(color online) 
Scatter plot of the overlaps between contacting particles at $\phi = 1.40$, and at $\phi = 1.55$.
Points below the diagonal in the top right quadrant correspond to overlaps whose magnitude is decreased on compression. The top left quadrant shows newly formed contacts, and the bottom right quadrant contains contacts that are destroyed. 
The figure clarifies that the peak at small overlaps in $P_Z(\delta)$ at $\phi = 1.55$ is mainly the result
of the relaxation of larger overlaps that existed at $\phi = 1.40$.
}
\end{figure}

These small overlaps must mainly result from the relaxation of old compressed bonds. Indeed,
since at $\phi = 1.40$ the overlap distribution close to $\delta = 0$ is small, $P_Z(\delta \simeq 0) \simeq 0$, 
we know that at this volume fraction there are very few particles that almost touch and so can be brought into contact on compression. 
The fact that the evolution of the overlap distribution on compression is characterized not only by the growth
of a peak at small overlaps but also by a decrease of the peak at large overlaps, further suggests that
some compressed bonds relax. To see this process at work we investigate the evolution
of the overlaps in a system prepared at $\phi = 1.40$ and then compressed quasistatically, with energy minimized at each step in volume fraction, to $\phi = 1.55$.
In the scatter plot of Fig.~\ref{fig:scatter} we show the overlaps $\left(\delta_{ij}^{(\phi = 1.40)},\delta_{ij}^{(\phi = 1.55)}\right)$ of each pair of particles that is in contact before and/or after the compression. 
The figure illustrates that a number of large overlaps at $\phi = 1.40$ become small at $\phi = 1.55$. While there also some new bonds with small overlaps, the majority of small overlaps arises by rearrangement of previously strongly compressed bonds.
This clarifies that the formation of a relatively small number of contacts with particles in the second coordination shell ($Z \simeq 8$ at $\phi = 1.55$)
induces a rearrangement of the whole force network, with the latter leading to an abundance of small contacts.

\subsection{Avalanches}
We saw in Fig.~\ref{fig:scatter} that, when a system beyond the first jamming crossover is compressed in a quasistastic manner, particle overlaps change significantly, sometimes by amounts of the order of the overlap itself.
This suggests that these changes do not originate from a continuous deformation of the system,
but rather from large scale catastrophic rearrangements, i.e.\ avalanches related to mechanical instabilities.
Indeed, on increasing the volume fraction one changes 
the linear system of equations that governs the mechanical equilibrium of the system, 
and may therefore induce a mechanical instability. 
These instabilities are well known to occur in athermal systems under quasistatic shear,
where they lead to sudden drops of the shear stress. 
To our knowledge avalanches in compression,
which are expected to lead to discontinuities in the pressure,
have never been reported. 

\begin{figure}[!!t]
\includegraphics*[scale=0.33]{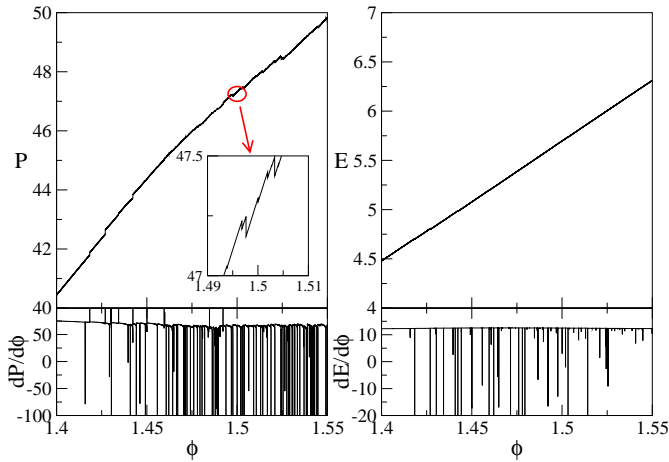}
\caption{\label{fig:compression} (color online) 
Evolution of the pressure (left panels) and of the energy (right panels)
of a system compressed quasistatically from $\phi = 1.40$ to $\phi = 1.55$.
The lower panels show the corresponding derivatives, evaluated with volume fraction steps of $d\phi = 10^{-5}$.
The figure demonstrates the existence of avalanches that cause a rearrangement of 
the force network of the system.
}
\end{figure}

We have confirmed this avalanche scenario by investigating the volume fraction dependence of the pressure, $P$, and of the energy, $E$,
of a system of $N = 10^3$ particles that is compressed quasistatically from $\phi = 1.40$ to $\phi = 1.50$ in 
increments of $\delta \phi = 10^{-5}$. Fig.~\ref{fig:compression} shows that the volume fraction dependence of the pressure 
is characterized by a large number of discontinuities; these are more clearly visible in the derivative $dP/d\phi$. Discontinuities are also present in the energy as a function of $\phi$. They are small and therefore difficult to see by eye on the scale of the graph, but can again be detected by looking at the 
volume fraction dependence of $dE/d\phi$.
Note that all avalanches induce a decrease of the elastic energy of the system, as expected.
On the other hand, even though most avalanches also induce a decrease in pressure, there are some that have the opposite effect.
Since on compressing the system the average value of the pressure continues to increase, the 
compression-induced dynamics clearly does not reach a steady state, providing an important distinction to the case of quasistatic shear. 
In this respect, avalanches in compression bear more resemblance to those associated with Barkhausen noise in ferromagnetic systems.

\subsection{Ordering}
Although global ordering is not expected in bidisperse systems such as the one studied here, it is still possible that compression to high volume fractions may change the local order around particles.
We next show that such a change does indeed occur at the first jamming crossover. To do this we measure local ordering via an 
extension of the local bond-orientational order parameter probing $n$-fold symmetry
in two dimensions~\cite{Nelson79}. This extended parameter is defined as follows:
\begin{equation}
\Psi^{n}_{q} = \frac{1}{N}\sum_{i=1}^N\frac{1}{n_i^{(q)}}\left| \sum_{j = 1}^{n_i} \delta_{ij}^q
e^{ni\theta_{ij}}\right|;
\;\;
n_i^{(q)} = \sum_{j=1}^{n_i} \delta_{ij}^q.
\end{equation}
Here the first sum runs over all particles $i$, and the second one over all $n_i$ overlaps $\delta_{ij}$ of particle $i$ with other particles. The angle $\theta_{ij}$ measures the orientation of each bond.
When $q = 0$, all bonds contribute equally to $\Psi^n_q$, which then reduces to the commonly investigated
local ordering parameter~\cite{Nelson79}. When $q \neq 0$ the weight of a bond depends
on its magnitude: $\Psi^{n}_{q}$ is mainly affected by small
overlaps if $q < 0$, and by large ones if $q > 0$.

\begin{figure}[!!t]
\includegraphics*[scale=0.33]{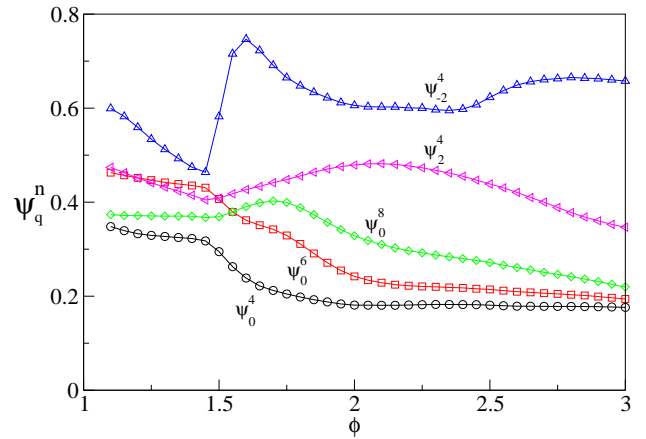}
\caption{\label{fig:ordering}
(color online) Volume fraction dependence of the order parameter $\Psi^n_q$ for different $n$ and $q$, as indicated.
}
\end{figure}

Considering that the average contact number per particle is $Z = 4$ at the jamming transition, $Z = 6$ before
the first jamming crossover, and that it reaches $Z = 8$ at $\phi \simeq 1.55$, 
we have investigated the volume fraction dependence of the usual ($q = 0$) bond order parameter for $n = 4$,$6$ and $8$. 
Fig.~\ref{fig:ordering} shows that $\Psi^4_0$ and $\Psi^6_0$ decrease monotonically with the volume fraction,
and that this decrease is fastest close to the first jamming crossover, at $\pj \simeq 1.40$. 
The order parameter $\Phi^8_0$, on the other hand, is non-monotonic in $\phi$, exhibiting a very small maximum close to $\pj \simeq 1.40$ as well as a more pronounced one at larger $\phi$.

This result is somewhat surprising, as particles with a local environment exhibiting an $8$-fold symmetry
are also characterized by a 4-fold symmetry, which one might expect to be revealed by $\Psi^4_0$.
To clarify this we investigate the influence of small and large overlaps on the order parameter,
by looking at $\Psi^4_{-2}$ and $\Psi^4_{2}$, respectively. The results for these parameters in Fig.~\ref{fig:ordering}, which show a clear peak in $\Psi^4_{-2}$,
demonstrate that at the jamming crossover small bonds arrange locally with 4-fold symmetry as in a square lattice.
There is a much broader peak in $\Psi^4_2$, suggesting that the first jamming crossover also leads to an increase of the square symmetry of strongly compressed bonds.

\begin{figure}[t!!]
\includegraphics*[width=7cm,height=7cm]{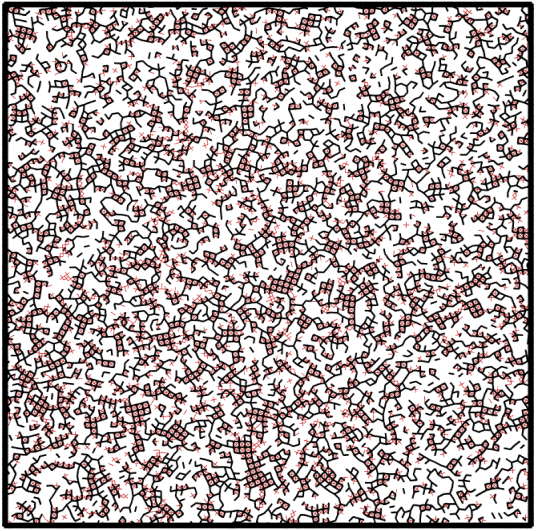}
\caption{\label{fig:fn} 
(color online) Partial contact network, for $\phi = 1.55$. Thick black 
and thin red segments correspond to overlaps greater that $\delta_{\rm max} = 0.24$
and smaller that $\delta_{\rm min} = 0.05$, respectively. 
}
\end{figure}

We explicitly illustrate the local square ordering of bonds induced by the jamming crossover, focussing on a
system with volume fraction $\phi = 1.55$. At this value of the volume fraction the overlap distribution has 
a minimum at $\delta_{\rm min} \simeq 0.05$, and a maximum at $\delta_{\rm max} \simeq 0.24$, 
as shown in Fig.~\ref{fig:Pd}. This suggests labelling an overlap $\delta$ as 
small if $\delta < \delta_{\rm min}$, and as large if $\delta > \delta_{\rm max}$. 
A visualization of the contact network restricted to such large and small overlaps, shown in Fig.~\ref{fig:fn}, clearly 
reveals the presence of local order. One locally identifies two interpenetrating square 
networks, at a $\pi/2$ angle, and made respectively of bonds with small and of large overlaps. 

\subsection{Softening}

\begin{figure}[t!!]
\includegraphics*[scale=0.33]{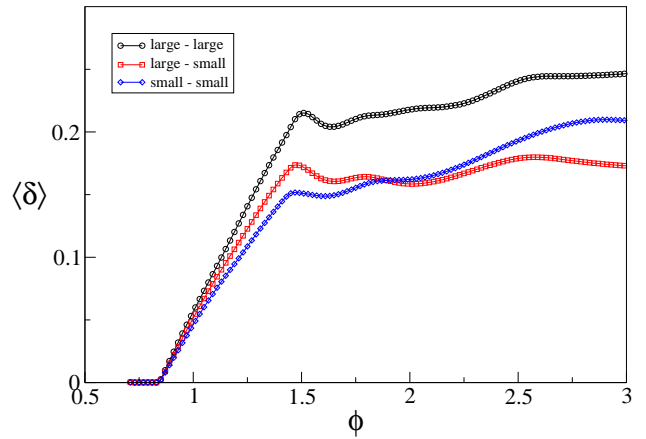}
\caption{\label{fig:overlap} (color online) 
Volume fraction dependence of the average value of the overlap between particles of different size.
}
\end{figure}

Next we look at the link between the non--affine rearrangement of the force network occurring at the jamming crossover and the observed softening of the system,
particularly the decrease of the bulk modulus. 
That there should be such a link can be seen by considering a monodisperse system of particles of diameter $D$
in two dimensions; the generalization to a discrete mixture or a genuinely polydisperse system is straightforward.
The bulk modulus is given by
\begin{equation}
 K = \phi \frac{dP}{d\phi} = \phi \frac{d}{d\phi}\left[ \frac{1}{2V} \sum_{c=1}^{N_c} \delta_c(D-\delta_c) \right],
\end{equation}
where the sum extends over all $N_c$ contacts, and $V$ is the volume of the system. 
Assuming small overlaps, $\delta_c \ll D$, this simplifies to
\begin{equation}
K \propto \phi \frac{d}{d\phi} Z \phi \<\delta\>,
\end{equation}
where $\<\delta\> = (1/N_c) \sum_c \delta_c$ is the overlap averaged over all contacts. 
To determine when the system softens, i.e.\ $dK/d\phi < 0$, we assume that the most important
effect of the jamming crossover is not the variation of the mean number of contacts, but 
rather the restructuring of the force network, and therefore approximate $Z$ as constant.
With this assumption one finds that
\begin{equation}
\frac{dK}{d\phi} < 0 \iff \<\delta\> + 3\frac{d\<\delta\>}{d\phi} + \phi \frac{d^2\<\delta\>}{d\phi^2} < 0.
\label{eq:softening}
\end{equation}
Accordingly, a system softens when the average overlap decreases quickly: the 
negative contribution of the second and possibly the third term in the above expression
may then overcome the positive contribution of the first term.
In the bidisperse case, the softening condition is analogous to Eq.~(\ref{eq:softening}), but involves
the average overlaps between the different particle species, large--large, large--small and small--small. 
Fig.~\ref{fig:overlap} shows that these average overlaps have a non monotonic volume
fraction dependence, and that their maximum values occur roughly at the jamming crossover. 
Accordingly, the rearrangement of the force
network at the jamming crossover, which leads to the formation of a large fraction of small contacts, 
explains via Eq.~(\ref{eq:softening}) the softening of the system.

We note as an aside that in the above approximation the pressure of the system scales as $P  \propto \phi \<\delta\>$.
A decrease of $\<\delta\>$ could therefore lead to an instability of the system 
if $d\<\delta\>/d \phi < -\<\delta\>/\phi$. We do not observe such an instability in our system.

\subsection{Normal modes}

One final perspective on the first jamming crossover is 
via the density of normal modes $D(\omega)$ and the spatial structure of the normal mode vectors. These
contain information
on the mechanical and the dynamical properties of the system. In systems
of particles interacting via repulsive potentials, $D(\omega)$ has been investigated mostly 
close to the jamming transition. At higher densities little is known about its variation with $\phi$
and about the spatial properties of the normal modes.

\begin{figure}[!t]
\includegraphics*[scale=0.55]{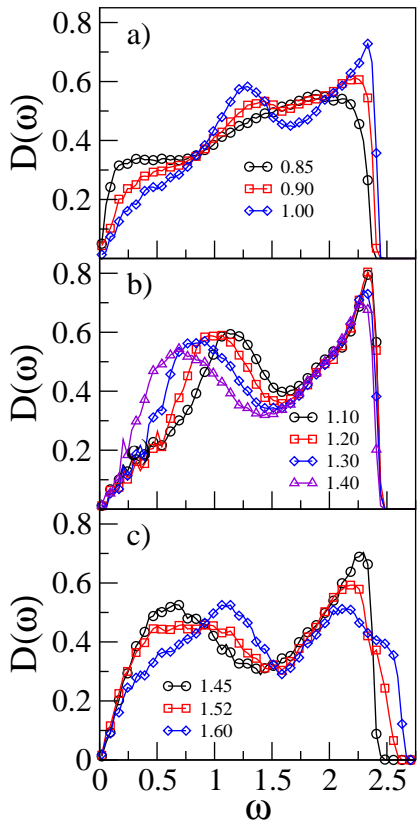}
\caption{\label{fig:domega} 
(color online) 
Evolution of the density of states with volume fraction close to the jamming transition (a), between the jamming transition
and the first jamming crossover (b), and around the jamming crossover (c).
}
\end{figure}

Here we describe the evolution of the density of states $D(\omega)$ in a volume fraction range extending from the jamming
transition to beyond the first jamming crossover. This evolution is characterized
by three different regimes, illustrated in the three panels of Fig.~\ref{fig:domega}:
changes occurring close to the jamming transition, between the jamming transition and the first jamming
crossover, and above the first jamming crossover.

The density of states for $\phi = 0.85$ (Fig.~\ref{fig:domega}a), shows features typical of systems 
close to the jamming transition. Debye scaling $D(\omega) \propto \omega$ is obeyed up to a characteristic frequency $\omega^*$. 
At higher frequencies $D(\omega)$ is first constant, and then grows towards a maximum value. 
If we now increase the volume fraction, $\omega^*$ increases and the plateau disappears. 
In addition, a new peak emerges, so that $D(\omega)$ becomes characterized by two peaks. 
Fig.~\ref{fig:domega}b illustrates that, on further increasing $\phi$, $D(\omega)$ mainly evolves via a
shift of the first peak towards smaller frequencies, while decreasing in height. 
This is in line with the approach to the first jamming crossover: the characteristic frequency (of the first peak) decreases again, 
making the system similar to one near the conventional jamming transition.
Finally, above the first jamming crossover, the first peak decreases in height as we move away from the first jamming crossover. 
A new peak emerges at a slightly larger frequency, as in Fig.~\ref{fig:domega}c, as a precursor to the approach to the second jamming crossover.

Next we wish to understand the spatial structure of the normal mode vectors.
Close to the jamming transition, conventional parameters such as the participation ratio and the bond stretching parameters
have not provided a clear answer to this question~\cite{Silbert2009,Zeravcic2008,Xu2010}. 
Here we show that some useful information can be extracted
at higher volume fractions from an extended bond--stretching parameter that
is designed to probe the behavior of overlaps of a given magnitude. 
For a generic normal mode $a$ and for a typical overlap $\delta$, we compute
\begin{equation}
S_a(\delta) = \left[ \frac{\sum_{\<i,j\>} \left| ({\bf u}_i^a - {\bf u}_j^a) {\bf n}_{ij} \right|^2 \Delta(\delta-\delta_{ij})}{\sum_{\<i,j\>} \left| {\bf u}_i^a - {\bf u}_j^a\right|^2 \Delta(\delta-\delta_{ij}) } \right]^{1/2},
\end{equation}
where $\Delta(x)$ indicates the Dirac $\delta$--function. 
Here ${\bf u}_i^a$ is the polarization vector of particle $i$ in mode $a$, 
${\bf n}_{ij} = ({\bf r}_i-{\bf r}_j)/r_{ij}$, and the sums run over all contacts.
The usual bond stretching parameter is as above, but lacks the delta functions and is therefore $\delta$ independent.
The contribution to $S_a(\delta)$ of a bond $\delta_{ij}$ of magnitude $\delta$ is close to unity when the bond is stretched or compressed, 
while conversely it is close to zero when the bond rotates. Accordingly,
values of $S_a(\delta)$ close to one indicate that all bonds of typical size $\delta$ stretch or compress, 
while values close to zero tell us than the majority of the bonds rotates.
In the actual calculation of $S_a(\delta)$, data are averaged over a small range of overlaps $\delta$ and over normal modes with frequencies in a range around some specified $\omega$. 
We re-emphasize that while the conventional bond--stretching parameter characterizes the deformation of the whole contact network of mode $a$,
$S_a(\delta)$ characterizes the deformation of contacts with overlaps of magnitude $\delta$. 

\begin{figure}[t!!!]
\includegraphics*[scale=0.35]{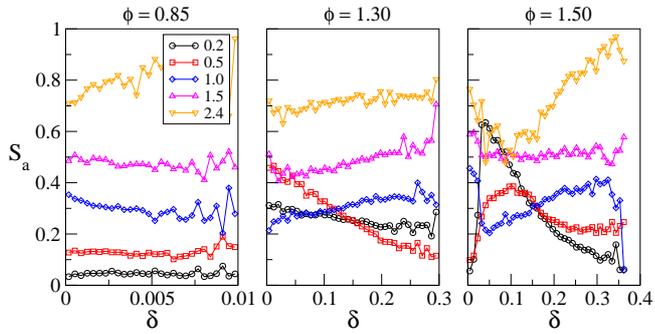}
\caption{\label{fig:Sa} 
(color online) 
Extended bond stretching parameter $S_a(\delta)$, as a function of the typical overlap, for different frequencies shown in the legend.
Each panel refers to a different value of the volume fraction, as indicated.
}
\end{figure}

We show in Fig.~\ref{fig:Sa} graphs of $S_a(\delta)$ for different values of the normal mode frequency, with separate panels referring
to different values of the volume fraction as indicated.
One notes first that on increasing the frequency the typical value of $S_a(\delta)$ increases for all volume
fractions, indicating that at high frequencies most of the bonds stretch. 
At volume fraction $\phi = 0.85$, which is slightly above the jamming transition, 
$S_a(\delta)$ is roughly $\delta$ independent. This indicates
that small and large bonds play a similar role in the deformation of the system.
This observation can be rationalized at least qualitatively from the fact that at such relatively moderate volume fractions the overlap probability distribution
decays exponentially. Accordingly, there is no strong separation between large and small overlaps, and in addition
the number of large overlaps is small. 

At $\phi = 1.30$, in between the jamming transition and the first jamming crossover, $S_a(\delta)$ depends more noticeably on the value of the overlap $\delta$.
In particular, at low frequencies, $S_a$ decreases with increasing $\delta$. Accordingly, low frequency modes are characterized
by a stretching or compression of bonds with small overlaps, and by the rotation of the larger ones. This is in agreement with the observation that
the energy cost of compressing a bond is lower for bonds with small overlaps than with large ones. The $\delta$ dependence of $S_a$ then
changes on increasing $\omega$, so that at high frequencies bonds with large overlaps stretch more.

By the time we reach volume fraction $\phi = 1.50$, above the first jamming crossover, the form of $S_a$ has changed significantly,
particularly at small frequencies. For instance, at $\omega = 0.2$, $S_a(\delta)$ is characterized by a high peak,
and assumes small values both for small and large overlaps. The peak occurs at a value of the overlap close to the location of the minimum of $P_Z(\delta)$, $\delta_{\rm min} = 0.05$ (see Fig.~\ref{fig:Pd}).
The fact that $S_a$ assumes small values both at small and large values of $\delta$ may be linked to the fact that above the first jamming crossover small and large overlaps are spatially correlated, as suggested by Fig.~\ref{fig:fn}.
The presence of a peak of $S_a$ at $\delta = \delta_{\rm min}$ could therefore be a consequence of the low frequency modes being characterized by relative rotation of clusters of particles. Bonds within these clusters would have either quite small or rather large overlaps, and bond stretching would occur at bonds with intermediate overlaps that connect these clusters. The possible presence of counter-rotating clusters of particles
is intriguingly reminiscent of the rigid unit modes which have been related to the density anomalies of network forming liquids~\cite{Dove}.

\begin{figure}[t!!!]
\includegraphics*[scale=0.25]{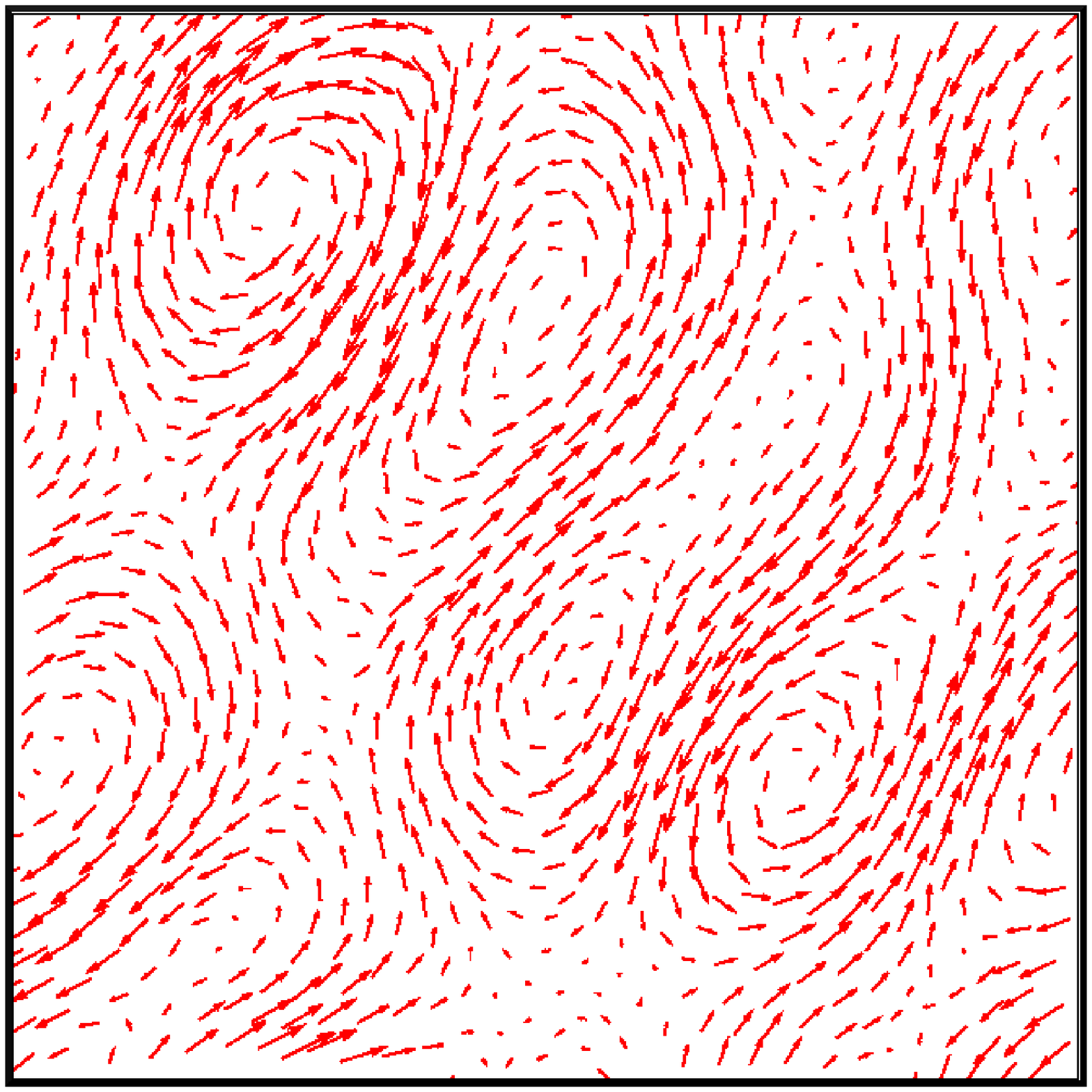}
\includegraphics*[scale=0.25]{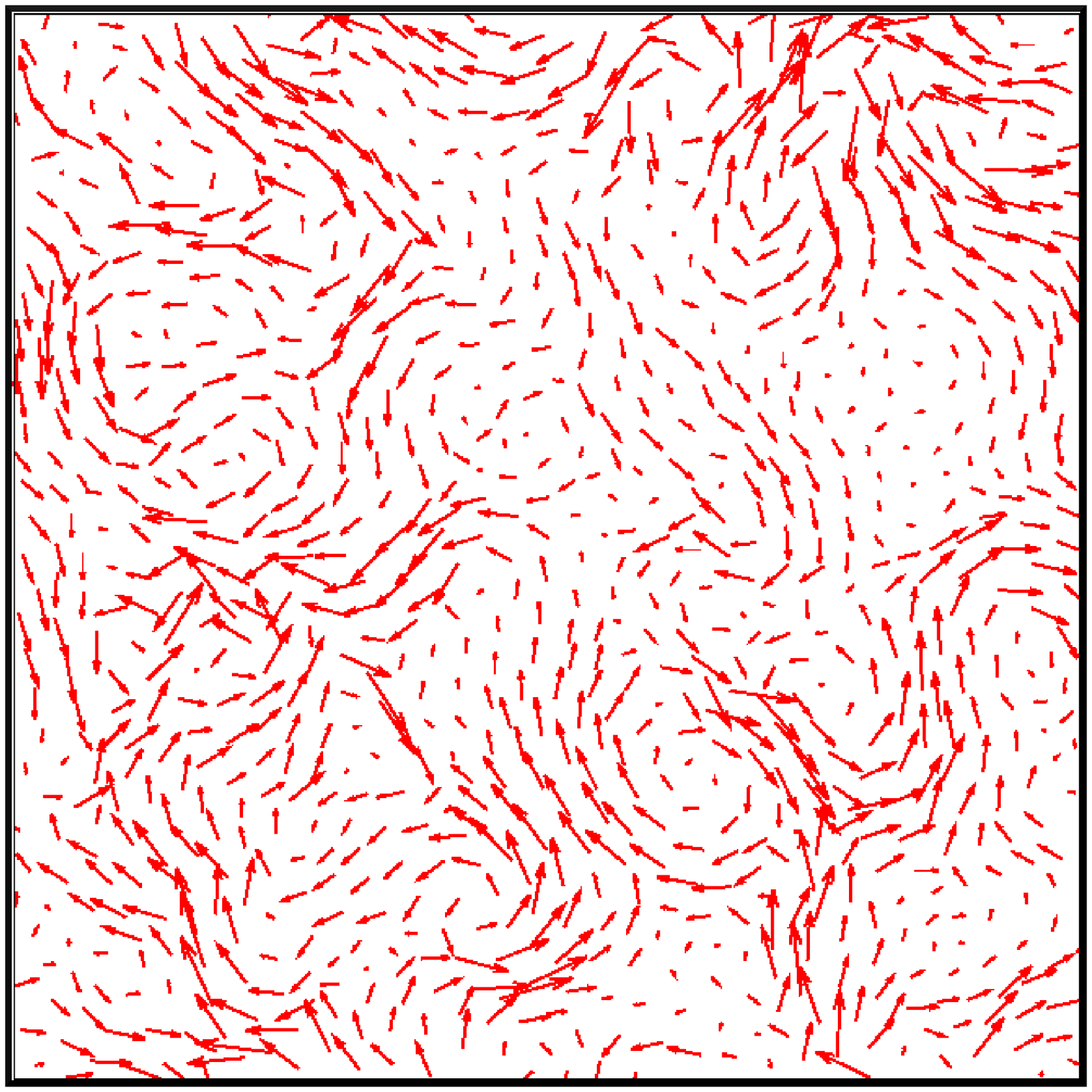}
\caption{\label{fig:eigenvector} 
(color online) 
Typical eigenvector of a $N = 10^3$ particle system at $\omega = 0.2$, and $\phi = 1.3$ (left) and $\phi = 1.5$ (right).
}
\end{figure}

The changes of the spatial properties of the low frequency energy modes as we go beyond
the first jamming crossover in volume fraction can also be appreciated by direct visualization. 
As illustrated in Fig.~\ref{fig:eigenvector}, below the crossover low frequency eigenvectors have 
a large scale swirling structure, while above the crossover they look more disordered. 

\section{Conclusions} 
We have shown that the first jamming crossover of a binary mixture of disks interacting
via a repulsive harmonic potential occurs at a well defined volume fraction, and that it
induces well defined structural changes. 
These changes are related to global rearrangements following plastic instabilities,
and lead to the formation of an abundance of small overlaps.
In turn, the large number of small overlaps is responsible for a decrease in the average value
of the overlap, and hence for the softening of the system. 
We have characterized the geometric and mechanical changes occurring at the crossover using
order parameters designed specifically to probe bonds with different overlaps and hence contact forces.
In particular, the study of the spatial structure of the low frequency eigenvectors suggested
that they may share some features of rigid unit modes~\cite{Dove}. Accordingly, these modes should play an important role in causing the density anomalies of these systems~\cite{toappear}.
It would therefore be interesting to repeat our zero--temperature study in simple models of water and of some
other liquids with density anomalies~\cite{Buldyrev2009}.

An intriguing question raised by our results concerns the connection between the geometric and mechanical properties of 
particle systems. Indeed, since connectivity strongly influences the mechanical
properties of a network~\cite{phillips}, then close to the jamming transition it has been possible to 
express the mechanical moduli in terms of the excess contact number $\Delta Z$~\cite{vanHecke2010,Liu2010}. 
These scaling relations predict that the bulk and the shear modulus both vary monotonically with $\Delta Z$.
Our results clarify that this monotonic behavior breaks down at jamming crossovers occuring at higher volume fractions, and therefore prove that the mechanical properties are not simply related to the connectivity.
Simple arguments to explain mechanical properties in terms of geometric ones away from jamming
should therefore be developed; we speculate that they would need to take account of the heterogeneity of the force network as a key feature of highly compressed particle systems.

\begin{acknowledgments}
MPC thanks the Dept.\ of Mathematics, King's College London, for hospitality,
and MIUR-FIRB RBFR081IUK for financial support.
\end{acknowledgments}


\begin{thebibliography}{40}
\bibitem{note}{
We have checked for some $\alpha$ and $\phi$ that we are working in a region of linear response. 
However, the derivatives we consider are smaller than the ones calculated in the regime of infinitesimal compressions of the system. 
Indeed, quantities calculated in this limit would not capture the presence of avalanche-like rearrangements on compression.
}
\end{thebibliography}
\end{document}